\documentclass[conference]{IEEEtran}
\IEEEoverridecommandlockouts
\usepackage{cite}
\usepackage{amsmath,amssymb,amsfonts}
\usepackage{algorithmic}
\usepackage{graphicx}
\usepackage{textcomp}
\usepackage{xcolor}
\def\BibTeX{{\rm B\kern-.05em{\sc i\kern-.025em b}\kern-.08em
    T\kern-.1667em\lower.7ex\hbox{E}\kern-.125emX}}

\makeatletter
\newcommand*\titleheader[1]{\gdef\@titleheader{#1}}
\AtBeginDocument{%
  \let\st@red@title\@title
  \def\@title{%
    \bgroup\normalfont\small\centering\@titleheader\par\egroup
    \vskip1.0em\st@red@title}
}
\makeatother

\usepackage{fancyhdr}
\newcommand*\titlefooter[1]{\gdef\@titlefooter{#1}}
\AtBeginDocument{%
  \fancyhf{} 
  \fancyfoot[C]{\@titlefooter}
  
  \fancypagestyle{firstpagestyle}{
    \fancyhf{} 
    \fancyfoot[L]{\footnotesize \@titlefooter} 
    
  }
  \pagestyle{plain}
}

\title{Toward Entailment Checking:\\ Explore Eigenmarking Search}

\makeatletter
\titleheader{Manuscript: 2025 3rd International Conference on Advancement in Computation \& Computer Technologies (InCACCT)}
\titlefooter{{\small XXX-X-XXXX-XXXX-X/XX/\$XX.00 \textcopyright20XX IEEE}}
\makeatother

\author{\IEEEauthorblockN{Tatpong Katanyukul}
\IEEEauthorblockA{\textit{Deparment of Computer Engineering} \\
\textit{Khon Kaen University}\\
Khon Kaen, Thailand \\
tatpong@kku.ac.th \\
ORCID 0000-0003-3586-475X}
}

\begin{document}

\maketitle

\thispagestyle{firstpagestyle}

\begin{abstract}
    Logic entailment is essential to reasoning,
    but entailment checking has the worst-case complexity of an exponential of the variable size.
    With recent development, quantum computing when mature
    may allow an effective approach for various combinatorial problems, 
    including entailment checking.
    Grover algorithm uses Grover operations, selective phase inversion and amplitude amplification
    to address a search over unstructured data with quadratic improvement from a classical method.
    Its original form is intended to a single-winner scenario: exactly one match is promised.
    Its extension to multiple-winner cases employs probabilistic control over 
    a number of applications of Grover operations, while a no-winner case is handled by time-out.
    
    Our study explores various schemes of ``eigenmarking'' approach.
    Still relying on Grover operations, but the approach introduces additional qubits 
    to tag the eigenstates.
    The tagged eigenstates are to facilitate an interpretation of the measured results and
    enhance identification of a no-winner case (related to no logic violation in entailment context).
           
    Our investigation experiments three variations of eigenmarking on a two-qubit system 
    using an IBM Aer simulator.
    The results show strong distinguishability in all schemes with the best relative distinguishabilities of 19 and 53 in worst case
    and in average case, respectively.
    Our findings reveal a viable quantum mechanism to differentiate a no-winner case 
    from other scenarios, which could play a pivot role in entailment checking and logic reasoning in general.
    
\end{abstract}


\section{Introduction}
Entailment is crucial for logic reasoning. 
It allows truth inference of an entailed sentence 
from its entailer.
However, in the worst-case\cite[pp. 240]{RN2022}, entailment checking is exponential of the number of logical symbols involved.

With a prospect on recent quantum computing development\cite{KimEtAl2023}, 
we explore an extension to Grover search\cite{Grover97} to address the entailment checking.

Grover search is used extensively in many quantum algorithms\cite{Ambainis04}.
It addresses a search over unstructured data with quadratic improvement\cite{BBHT96} 
from a classical method.
Its speed up is achieved through quantum mechanical properties such as superposition and entanglement,
which together allow computational parallism. 
Its key mechanism employs selective phase inversion and inversion about mean.
These two operations are often jointly called ``Grover operation,''
designed to amplify the amplitude (and consequently probability) of the answer for that we are searching.

Grover search in its original form is intended to a single-winner scenario: exactly one answer is promised.
Its extension\cite{BBHT96} to multiple-winner cases employs probabilistic control over 
a number of applications of Grover operations, while a no-winner case is handled by time-out.

Entailment checking is distinct from a general search problem 
in that task is mainly to distinguish a no-winner case (i.e., no entailment violation) 
from other cases (i.e., there is at least some violation).
To address this issue efficiently, we explore an approach using additional qubit(s),
facilitating handling of various cases from no-winner, one-winner, multiple-winner all the way to all-winner scenarios.
This approach is collectively called ``eigenmarking.''
Three variations are studied, particularly for differentiation between no-winner case and other cases.
Although eigenmarking is intended for entailment checking, its applicability could go beyond this goal 
and may in part advance the development of quantum search, its applications, and quantum algorithms in general. 

\section{Background}

\paragraph{Entailment}

Sentence (or knowledge base) $\alpha$ entails sentence $\beta$,
written with notation $\alpha \vDash \beta$,
means that every model%
\footnote{%
In logical reasoning context, 
a model means a specific set of truth values for all logical symbols in both sentences.%
}
that is true in the entailer $\alpha$ must be true in the entailee $\beta$.
There are two main approaches for entailment checking\cite{RN2022}:
\emph{theorem proving}
and \emph{model checking}.
Theorem proving applies rules of inference to the entailer to eventually derive the entailee.
Model checking goes through combinations of truth values for all logical symbols
to validate the relation of an entailer and its entailee, i.e.,
there is no model that is true in $\alpha$ but false in $\beta$.
It is equivalent to test unsatisfiability of $\alpha \wedge \neg \beta$.




\paragraph{Quantum computation}
Exploiting quantum mechanical properties in computation is the main characteristic of quantum computation.
Quantum mechanical system evolves in two distinct modes\cite{NC2016, YM2008}.
(1) Linear evolution mode, its state $| \psi \rangle$ evolves according to Schr{\"o}dinger equation (SE)
based on Hamiltonian $\hat{H}$ (energy operator) of the system.
I.e.,
$i \hbar \frac{\partial | \psi \rangle}{\partial t} =  \hat{H} | \psi \rangle$
where $i = \sqrt{-1}$ and $\hbar$ is a reduced Planck constant.
This is equivalent to \textit{unitary transformation}, where the unitary operator corresponds to the Hamiltonian.
I.e.,
the solution to SE,
$| \psi \rangle = | \psi_0 \rangle \exp(-\frac{i}{\hbar} \hat{H} t)$
where $| \psi_0 \rangle$ is an initial state.
Let unitary operator $U \equiv \exp(-\frac{i}{\hbar} \hat{H} t)$ of specific time $t$, we have
$| \psi \rangle = U | \psi_0 \rangle$.

Quantum state $| \psi \rangle$ can be represented
in a summation form, e.g., a one-qubit state $| \psi \rangle = \alpha | 0 \rangle + \beta | 1 \rangle$,
or in a matrix form, e.g., $| \psi \rangle = [\alpha , \beta ]^T$ 
(the eigenstates $| 0 \rangle$ and $| 1 \rangle$ are implicit),
where $\alpha$ and $\beta$ are (probability) amplitudes.

With precise control on Hamiltonian, at least in theory, we can have
any unitary operator we want.
There are common unitary operators. 
E.g., not operator,
$X = \begin{bmatrix}0 & 1 \\ 1 & 0\end{bmatrix}$.
Hadamard operator, 
$H = \frac{1}{\sqrt{2}}\begin{bmatrix}1 & 1\\1 & -1\end{bmatrix}$.


      
(2) Measurement mode, its state collapses to one of its eigenstates upon measurement.
The eigenstates correspond to the measurement operator, which is applied to measure the system.
A probability of an eigenstate to which the state collapses
is a squared modulus of the amplitude of that eigenstate in the state before the measurement.
E.g., given a state $|\psi \rangle = \frac{1}{\sqrt{2}}|0\rangle + \frac{1}{\sqrt{2}}|1 \rangle$,
after the measurement in a standard basis,
the state after measurement $|\psi'\rangle$ has probability of 
$|\frac{1}{\sqrt{2}}|^2 = 0.5$ 
to be in $|0\rangle$ and similarly probability of $0.5$ to be in $|1\rangle$.

A quantum computer can be implemented using many technologies, e.g., superconducting circuit\cite{KKYOGO19},
and to control qubits can be done by changing Hamiltonian (consequently unitary transformation, $U$),
e.g., through microwave pulses\cite{KKYOGO19}.

\section{Grover algorithm}
Grover algorithm addresses a search problem:
given a function $f: \{0, 1\}^n \rightarrow \{0, 1\}$
with a promise that
there is exactly one answer $\mathbf{x}'$ 
such that
$f(\mathbf{x}) = 1$ if $\mathbf{x} = \mathbf{x}'$ and $f(\mathbf{x}) = 0$ otherwise.
A binary input $\mathbf{x}$ has $n$ qubits.
Thus, there are $N = 2^n$ possible combinations to search.

Suppose 
the underlying function is given as a corresponding unitary operator $U_f$, 
which
takes $n + 1$ qubits ($n$ for input $\mathbf{x}$ and $1$ for ancillary $y$)
and rotates $y$ qubit by $\pi$ (phase inversion) only when $f(\mathbf{x}) = 1$.
This could be done by phase kickback\cite{YM2008}, 
but for simplicity here we formulate it as a direct controlled phase rotation, i.e.,
$U_f \equiv |\mathbf{x}'\rangle \langle \mathbf{x}'| \otimes R_z(\pi)  + \sum_{\mathbf{x} \in \tilde{X}} |\mathbf{x}\rangle \langle \mathbf{x}| \otimes I$
where $\tilde{X}$ is a set of non-winning states;
identity $I = \begin{bmatrix} 1 & 0 \\ 0 & 1\end{bmatrix}$
and rotation $R_z(\theta) = \begin{bmatrix} 1 & 0 \\ 0 & e^{i \theta}\end{bmatrix}$.

\begin{figure}[htbp]
    \center{\includegraphics[width=0.5\textwidth]{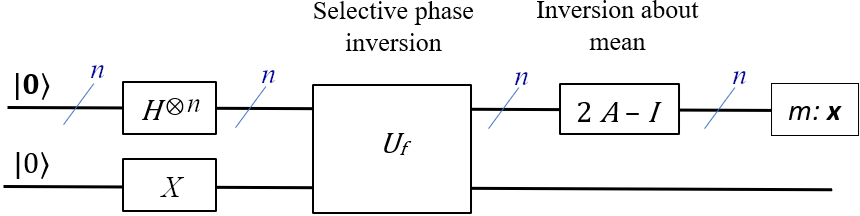}}
\caption{Original Grover search.}
\label{fig: grover}
\end{figure}

Grover algorithm (Fig.~\ref{fig: grover}) is as follows.
\begin{itemize}
    \item (1) Prepare $|\mathbf{x} \rangle$ in a ground state, 
          i.e., $|\mathbf{x}_0 \rangle = |\mathbf{0}\rangle \equiv |\underbrace{00\ldots0}_n \rangle$. 
    \item (2) Apply $H^{\otimes n}$,
            where $H^{\otimes n} = \underbrace{H \otimes H \otimes \cdots \otimes H}_{n}$ 
            is a tensor product of $H$.
            It is equivalent to apply $H$ independently to each qubit of $| \mathbf{x}_0 \rangle$,
            i.e.,
            $|\mathbf{x}_1\rangle = H^{\otimes n} |\mathbf{x}_0\rangle$.
    \item (3) Apply the phase inversion, 
            i.e., 
            $|\psi_2\rangle = U_f |\mathbf{x}_1, 1 \rangle$.
    \item (4) Apply the inversion about the mean to the input part,
            i.e.,
            segment $|\mathbf{x}_2 \rangle |y \rangle = | \psi_2 \rangle$
            and have
            $|\mathbf{x}_3 \rangle = (2 A - I) |\mathbf{x}_2 \rangle$,
            where $I$ is an identity matrix and an average matrix 
            $A = \frac{1}{N} \begin{bmatrix}
            1 & 1 & \ldots & 1 \\
            1 & 1 & \ldots & 1\\
            \vdots & \vdots & \ddots & \vdots \\
            1 & 1 & \ldots & 1
            \end{bmatrix}$.
    \item (5) Grover operation (steps 3 and 4) should be applied for $J$ time(s),
            where%
            \footnote{%
            Derived from \cite{BBHT96}.
            }%
             $J = \arg\min_j \frac{1}{N - 1} \cos^2 \left( (2j + 1) \sin^{-1}\sqrt{\frac{1}{N}} \right)$
             or simply $J = \mathrm{round}\left(\frac{\pi}{4} \sqrt{N} - \frac{1}{2}\right)$ when $N$ is large.
    \item (6) Measure the qubits.
\end{itemize}

The answer, which is the single winning state, 
then has the highest probability and the measured state is very likely to be the sought-after answer $\mathbf{x}'$.


For example, a case of $n = 2$ and the answer $\mathbf{x}' = \mbox{01}$, Grover search can be conducted as follows.

\begin{itemize}
    \item (1) Prepare $|\mathbf{x}_0 \rangle = |00 \rangle = [1, 0, 0, 0]^T$.
    \item (2) Apply Hadamard operator $|\mathbf{x}_1\rangle = H^{\otimes 2} |\mathbf{x}_0\rangle = [\frac{1}{2},\frac{1}{2},\frac{1}{2},\frac{1}{2}]^T$.
    \item (3) Apply the phase inversion, 
            i.e., 
            $|\psi_2\rangle = U_f |\mathbf{x}_1, 1 \rangle = U_f [0,\frac{1}{2},0,\frac{1}{2},0,\frac{1}{2},0,\frac{1}{2}]^T$
            $ = [0,\frac{1}{2},0,-\frac{1}{2},0,\frac{1}{2},0,\frac{1}{2}]^T$.
    \item (4) Apply the inversion about the mean to the input part,
            i.e.,
            segment $|\mathbf{x}_2 \rangle = [\frac{1}{2},-\frac{1}{2},\frac{1}{2},\frac{1}{2}]^T$ (and $|y\rangle = [0,1]^T$)
            and 
            $|\mathbf{x}_3 \rangle = (2 A - I) |\mathbf{x}_2 \rangle = [0,1,0,0]^T$.
    \item (5) $J = \mathrm{round}\left(\frac{\pi}{4} \sqrt{4} - \frac{1}{2}\right) = 1$.
    \item (6) Here, when measured, $|01\rangle$ will be observed with probability of $1$.
\end{itemize}

As analyzed by \cite{BBHT96},
the state can be written in term of the single answer (winning state) $|\mathbf{x}'\rangle$ and any non-answer $|\mathbf{x}\rangle$ as,

\begin{align}
    |\mathbf{x}_1\rangle &= H^{\otimes n} |\mathbf{x}_0\rangle 
    \nonumber \\
    &= \frac{1}{\sqrt{2^n}} \left( |\mathbf{x}' \rangle  
       + \sum_{\mathbf{x} \in \tilde{X}} |\mathbf{x} \rangle \right)
    \\
    |\psi_2\rangle &= U_f | \mathbf{x}, 1 \rangle
    \nonumber \\
    &= \frac{1}{\sqrt{2^n}} \left( -|\mathbf{x}', 1 \rangle  
       + \sum_{\mathbf{x} \in \tilde{X}} |\mathbf{x}, 1 \rangle \right)
    \nonumber \\
    &= \underbrace{\frac{1}{\sqrt{2^n}} \left( -|\mathbf{x}' \rangle 
       + \sum_{\mathbf{x} \in \tilde{X}} |\mathbf{x} \rangle \right)}_{|\mathbf{x}_2\rangle} \underbrace{|1 \rangle}_{|y\rangle}
    \nonumber \\
\end{align}
    
    Let $-k$ and $\ell$ be the probability amplitudes of the winning state $|\mathbf{x}' \rangle$ 
    and any non-winning state $|\mathbf{x} \rangle$ respectively.
    Therefore,  
    \begin{align}
    | \mathbf{x}_2 \rangle &=  \underbrace{-\frac{1}{\sqrt{2^n}}}_{-k} |\mathbf{x}' \rangle 
    + \underbrace{\frac{1}{\sqrt{2^n}}}_{\ell} \sum_{\mathbf{x} \in \tilde{X}} |\mathbf{x} \rangle
    \nonumber \\
    &= -k |\mathbf{x}' \rangle  + \ell \sum_{\mathbf{x} \in \tilde{X}} |\mathbf{x} \rangle
    .
    \end{align}

Next, inversion about the mean
\begin{align}
| \mathbf{x}_3 \rangle &= (2 A - I) | \mathbf{x}_2 \rangle
\nonumber .
\end{align}

Recall that $A$ is to average the probability amplitudes:
there is only one winning state $\mathbf{x}'$ (with amplitude $-k$), 
but there are $N - 1$ non-winning states (each with amplitude $\ell$), thus

\begin{align}
A | \mathbf{x}_2 \rangle &= \frac{- k + (N - 1) \ell }{N} \left( |\mathbf{x}' \rangle  + \sum_{\mathbf{x} \in \tilde{X}} |\mathbf{x} \rangle \right)
\\
2 A | \mathbf{x}_2 \rangle 
&= \frac{- 2 k + (2 N - 2) \ell }{N} 
\left( |\mathbf{x}' \rangle  + \sum_{\mathbf{x} \in \tilde{X}} |\mathbf{x} \rangle \right)
\\
(2 A - I) | \mathbf{x}_2 \rangle &= 
\left( \frac{- 2 k + (2 N - 2) \ell }{N} + k_2 \right) |\mathbf{x}' \rangle  
\nonumber \\
&\; + \left( \frac{- 2 k + (2 N - 2) \ell }{N} - \ell_2 \right) \sum_{\mathbf{x} \in \tilde{X}} |\mathbf{x} \rangle
\nonumber \\
&= \left( \frac{ (N - 2) k + (2 N - 2) \ell }{N} \right) |\mathbf{x}' \rangle  
\nonumber \\
&\; + \left( \frac{- 2 k + (N - 2) \ell }{N} \right) \sum_{\mathbf{x} \in \tilde{X}} |\mathbf{x} \rangle
\nonumber \\
\end{align}

That is, 
given state $|s\rangle = k | \mathbf{x}' \rangle + \ell | \mathbf{x} \rangle$,
Grover iteration (phase inversion and inversion about the mean) evolves the state to:

\begin{align}
| \hat{s} \rangle =& \left( \frac{N - 2}{N} k + \frac{2(N - 1)}{N} \ell \right) | \mathbf{x}' \rangle 
\nonumber \\
&\;
+ \left( - \frac{2}{N} k + \frac{N - 2}{N} \ell \right) \sum_{\mathbf{x} \in \tilde{X}} | \mathbf{x} \rangle
\end{align}

To analyze an optimal number of Grover operations, 
let denote $k_j$ and $\ell_j$ for amplitudes $k$ and $\ell$ after $j$ application(s).
For $N = 2^n$ and $n$ being a word length (a number of qubits in a solution state $|\mathbf{x} \rangle$),
$k_0 = \frac{1}{\sqrt{N}}$ and $\ell_0 = \frac{1}{\sqrt{N}}$
and
\begin{align}
k_j &= \frac{N - 2}{N} k_{j-1} + \frac{2(N - 1)}{N} \ell_{j-1}
\label{eq: kj} \\
\ell_j &= - \frac{2}{N} k_{j-1} + \frac{N - 2}{N} \ell_{j-1}
\label{eq: lj}
\end{align}
for $j = 1, 2, \ldots $.

\begin{figure}[htbp]
    \center{\includegraphics[width=0.5\textwidth]{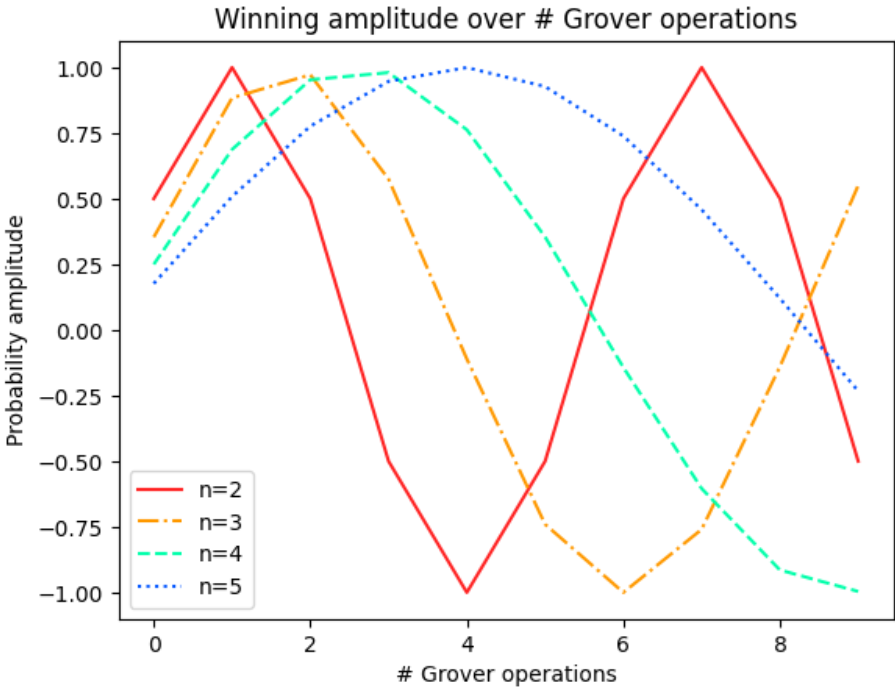}}
\caption{Evolution of a probability amplitude of the winning state per grover iterations at different numbers of qubits $N$.}
\label{fig: number of Grover operations}
\end{figure}

Figure~\ref{fig: number of Grover operations} shows 
how probability amplitudes of the winning state evolve 
over grover iterations under different settings.
Notice that their evolutions show strong sinusoidal patterns 
with frequencies directly associate to numbers of qubits.
These patterns are apparently realized in \cite{BBHT96},
\begin{align}
    k_j &= \sin \left((2j + 1) \theta \right)
\label{eq: kj closed}
    \\
    \ell_j &= \frac{1}{\sqrt{N -1}} \cos \left( (2j + 1) \theta \right)
\label{eq: ellj closed}
\end{align}
where $\theta = \sin^{-1} (\frac{1}{\sqrt{N}})$.
With equation~\ref{eq: ellj closed}, it is apparent 
that to minimize the chance of observing the non-answer state,
$\cos \sim 0 \Rightarrow \theta \sim \frac{\pi}{2}$, i.e., $(2 j^\ast + 1) \theta = \frac{\pi}{2}$
or $j^\ast \approx  \frac{\pi}{4 \theta} - \frac{1}{2} \approx \frac{\pi \sqrt{N}}{4} - \frac{1}{2}$ for a large $N$ (since $\theta = \sin^{-1} \frac{1}{\sqrt{N}} \approx \frac{1}{\sqrt{N}}$).

\section{Quantum eigenmarking}
Three eigenmarking schemes are explored here.
Despite the detail, they share the common mechanism: using additional qubit(s)
to facilitate differentiation particularly of a non-winner case from other cases.

\begin{figure}[htbp]
    \center{\includegraphics[width=0.5\textwidth]{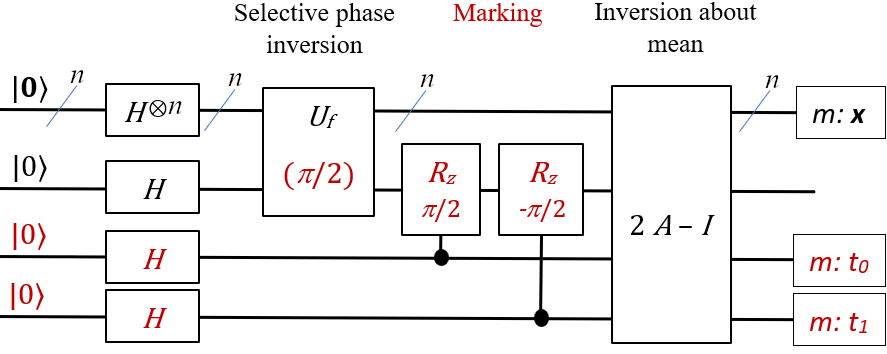}}
\caption{Grover search with eigenmarking. The averaging in operation $2 A - I$ is performed over every qubit, indicated by an enlarged block, c.f., one in original Grover search.}
\label{fig: eigenmarking}
\end{figure}

\begin{figure}[htbp]
    \center{\includegraphics[width=0.5\textwidth]{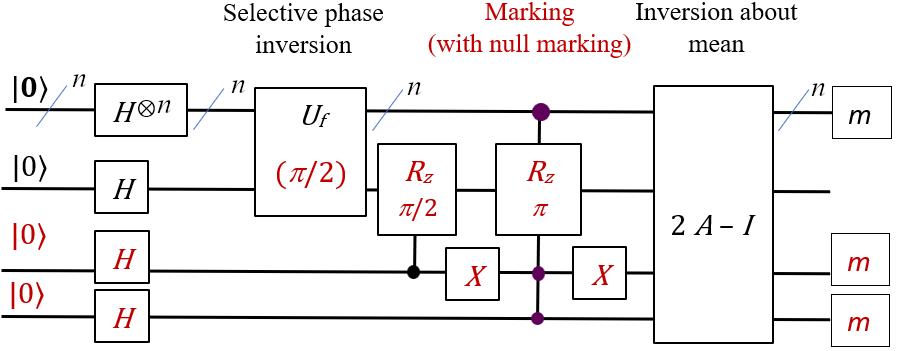}}
\caption{Eigenmarking with null marking.}
\label{fig: null marking}
\end{figure}

\begin{figure}[htbp]
    \center{\includegraphics[width=0.5\textwidth]{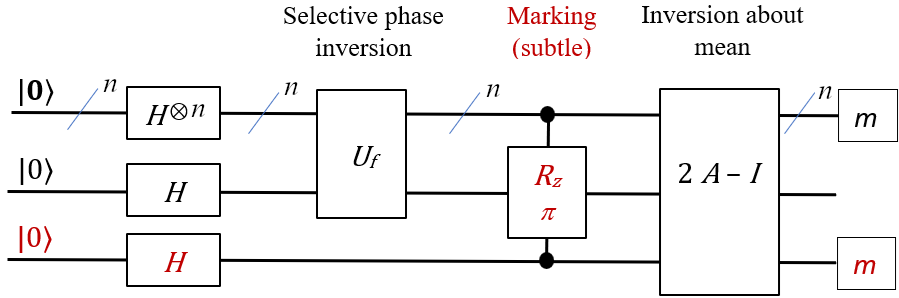}}
\caption{Eigenmarking with subtle marking.}
\label{fig: subtle marking}
\end{figure}

To be concise, while refering to the first scheme simply as eigenmarking (Fig.~\ref{fig: eigenmarking}), 
we will refer to eigenmarking with null marking (Fig.~\ref{fig: null marking}) and with subtle marking (Fig.~\ref{fig: subtle marking})
as null and subtle marking, respectively.

\paragraph{Eigenmarking}
This scheme modifies Grover search by (1) having two additional qubits $|t_1 t_0\rangle$ 
each has controlled-phase on the ancillary $|y\rangle$,
(2) having the $U_f$ shift phase by only $\pi/2$ (instead of full phase inversion, equivalent to $\pi$),
and (3) extending the inversion about the mean to average over all qubits (not just the input qubits).

Specifically,
\begin{itemize}
\item (1) have
$U_f \equiv |\mathbf{x}'\rangle \langle \mathbf{x}'| \otimes R_z(\pi/2) + \sum_{\mathbf{x} \in \tilde{X}} |\mathbf{x}\rangle \langle \mathbf{x}| \otimes I$;
\item (2) prepare
$|\psi_0\rangle = |\mathbf{x}, y \rangle = |\mathbf{0},0\rangle$;
\item (3) do
$|\psi_1\rangle = H^{\otimes (n+1)} |\psi_0\rangle$;
\item (4) apply Grover selection $|\psi_2\rangle = U_f |\psi_1\rangle$;
\item (5) apply marking:
let $|t_1, t_0\rangle = H^{\otimes 2} |0,0\rangle$
and segment $|\mathbf{x}_2\rangle | y_2 \rangle = |\psi_2\rangle$,
then $|t_0', y_3 \rangle = \mathbf{CR}(\pi/2) |t_0, y_2 \rangle$
and $|t_1', y_4 \rangle = \mathbf{CR}(-\pi/2) |t_1, y_3 \rangle$
where controlled phase rotation $\mathbf{CR}(\theta) = I \otimes |0\rangle \langle 0| + R_z(\theta) \otimes |1\rangle \langle 1|$;
\item (6) apply inversion about the mean to all qubits:
aggregate $| \psi_4 \rangle \equiv |t_1', t_0', \mathbf{x}_2, y_4 \rangle$
and $| \psi_5 \rangle = (2 A - I) | \psi_4 \rangle$;
\item (7) measure the qubits.

\end{itemize}

The rationale behind this design is that 
(1) with two additional qubits, the number of states will be quadruple:
even with a scenario of all winning states, the winning states are still minority 
and their amplitudes would be effectively amplified by inversion about the mean.
(2) With phase rotation by prefix, there will always be some states stand out, 
even in a scenario of no winning states.
Those stand-out states are amplified only by their prefix,
therefore they always show up roughly in unity and should be easily identified.

\paragraph{Null marking}
Similar to the first scheme, but picked a state to represent a null or no-winner scenario,
it uses phase rotation with multiple controls along with Not operators to select state $|t_1,t_0, \mathbf{x}\rangle$
$=|10 \underbrace{1\ldots1}_\mathbf{x}\rangle$ as a null state.

\paragraph{Subtle marking}
This scheme is more like a minimalist version of the null marking: 
only one additional qubit is employed 
and no modification to the $U_f$ 
(full phase inversion, just as in the original Grover algorithm).

Specifically, the evolution is as follows.
\begin{itemize}
\item (1) Prepare system in the ground state: $|\psi_0 \rangle = | \mathbf{0} \rangle$,
where $|\psi_0 \rangle = |t_0, \mathbf{x}, y \rangle$.
\item (2) Apply Hadamard: $|\psi_1 \rangle = H^{\otimes n+2} |\psi_0 \rangle$.
\item (3) Apply Grover selection: $|\psi_2 \rangle = (I \otimes U_f) |\psi_1 \rangle$.
\item (4) Apply marking: $|\psi_3 \rangle = \mathrm{MCR}(\pi)|\psi_2 \rangle$,
where multiple-qubit control phase $\mathrm{MCR}(\theta) = | 1,\mathbf{1} \rangle \langle 1,\mathbf{1}| \otimes R_z(\theta)  + \sum_{\mathbf{x} \neq | 1, \mathbf{1} \rangle} | \mathbf{x} \rangle \langle \mathbf{x}| \otimes I$.
\item (5) Do inversion about the mean: $|\psi_4 \rangle = (2 A - I) |\psi_3 \rangle$.
\item (6) Measure the qubits.

\end{itemize}



\section{Methodology}

\begin{figure}[htbp]
    \center{\includegraphics[width=0.48\textwidth]{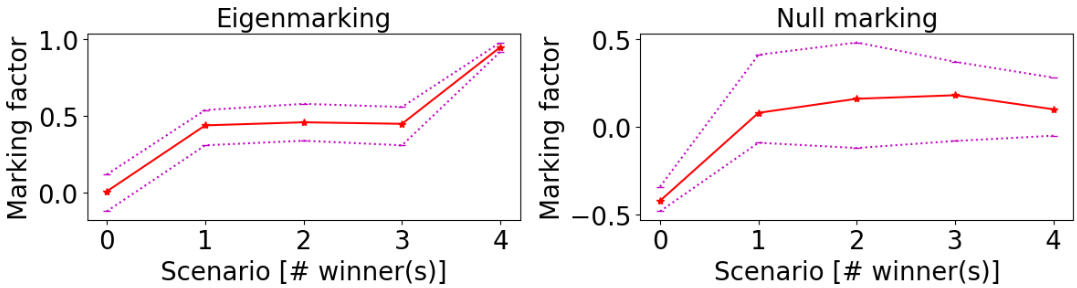}}
\caption{Marking factors using eigenmarking (left) and null marking (right).}
\label{fig: no win vs some eigen null}
\end{figure}

The three eigenmarking schemes are tested on all possible scenarios of a two-qubit system
using Qiskit (version 1.1.1) and Qiskit Aer Simulator (version 0.15.1).
Each treatment is repeated for 40 times, while each time the simulator simulates it for 1024 runs.
The main results are presented in Tables~\ref{tab: lambda0 vs others}
and~\ref{tab: win margins}
with visualization in Fig.~\ref{fig: no win vs some eigen null} and~\ref{fig: no win vs some subtle and win margin}.
All the detailed experimental results are shown in the Appendix.

\paragraph{Marking factor}

\begin{align}
M = \frac{w - w_0}{w + w_0}
\label{eq: marking factor} ,
\end{align}

where the definitions of $w$ and $w_0$ 
depend on the marking scheme.

For eigenmarking, 
$w = \max C_\text{01}$; 
$w_0 = \max C_\text{10}$
and $C_\text{xx}$ is a count of measured states with prefix \texttt{xx}.
The prefix \texttt{01} marks the answers,
while \texttt{10} marks complementary states.
The maximization is performed over states with the specified prefix.
E.g., 
$C_\text{01} = [84, 192, 71, 69]$ and $\max C_\text{01} = 192$
when
the measured results are
$\{
\ldots ,
\underbrace{84}_\text{`01 00'}, 
\underbrace{192}_\text{`01 01'},
\underbrace{71}_\text{`01 10'}, 
\underbrace{69}_\text{`01 11'}, 
\ldots  
\}$ 
 from 1024 runs.
 
For null marking, 
 $w = \max C_\text{01}$ (same as eigenmarking)
 and 
 $w_0 = C_\text{1011}$,
 which is simply the count of measurement on state \texttt{1011} (the null marking state).

For subtle marking,
 $w = \max C_\text{0}$, 
 which is the maximum count among prefix \texttt{0} (answer marking prefix).
 E.g., for two-qubit input,
 $C_0$ is a set of counts of state \texttt{000} to state \texttt{011}.
 The null count, $w_0 = C_\text{111}$,
 is the count of measurement on state \texttt{111} (the null marking state for this scheme).

The larger difference in marking factors between no-winner case and others
indicates the effectiveness of the corresponding scheme.

\begin{table}[htbp]
    \caption{Marking factor: mean(standard deviation)}
    \begin{center}
    \begin{tabular}{|l|c@{ }|c@{ }|c@{ }|c@{ }|c@{ }|}
    \hline
     & \multicolumn{5}{|c|}{\textbf{Scenario}: \# winner(s)} \\
    \cline{2-6}
    \textbf{Scheme} & \textbf{0} & \textbf{1} & \textbf{2} & \textbf{3} & \textbf{4} \\
    \hline
    Eigen.  &  0.01(0.05)  & 0.44(0.05)  & 0.46(0.05) & 0.45(0.05) & 0.95(0.01) \\
    Null.   & -0.42(0.04)  & 0.08(0.14)  & 0.16(0.15) & 0.18(0.10) & 0.10(0.07) \\
    Subtle. & -0.73(0.02)  & 0.19(0.33)  & 0.48(0.47) & 0.70(0.39) & 0.33(0.06) \\
    \hline
    \end{tabular}
    \label{tab: lambda0 vs others}
    \end{center}
\end{table}

\paragraph{Winning margin}
relative winning margin
is the relative difference between minimal count of the winning state and maximal count of the non-winning state.
We explore two versions: global version whose non-winning state can be any non-winning state
and local version whose candiates of the non-winning states are taken only those with the answer prefix.
Specifically, 
\begin{align}
W = \frac{c - c'}{c'}
\label{eq: winning margin} ,
\end{align}
where 
$c = \min_x C_x$ s.t. $x \in \Omega_w$;
$\Omega_w$ is a set of winning states;
and
$c' = \max_x C_x$ s.t. $x \not \in \Omega_w$
for a global margin
or
$c' = \max_{x \in \Omega_p} C_x$ s.t. $x \not \in \Omega_w$
for a local margin,
where $\Omega_p$ is a set of states with the answer prefix, i.e.,
prefix \texttt{01} for eigenmarking and null marking
and prefix \texttt{0} for subtle marking.
Note that no-winner and all-winner scenarios are excluded from computing
the winning margins, since these scenarios have no counts of either winner or non-winner states. 

\begin{figure}[htbp]
    \center{\includegraphics[width=0.48\textwidth]{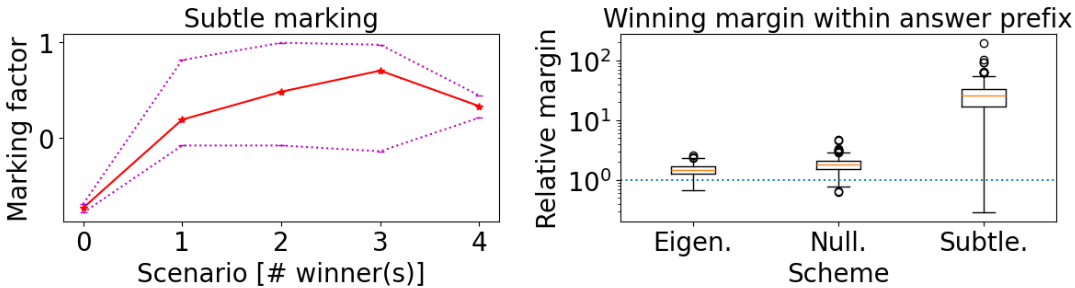}}
\caption{Marking factors using subtle marking (left) and winning margins (right).}
\label{fig: no win vs some subtle and win margin}
\end{figure}

\begin{table}[htbp]
    \caption{Winning margin: [min., mean, max.](std. dev.)}
    \begin{center}
    \begin{tabular}{|l|c|c|c|}
    \hline
    \textbf{Scheme} &  \multicolumn{2}{|c|}{\textbf{Relative winning margin}} & \textbf{Local} \\
    \cline{2-3}
                    & \textbf{Global}   & \textbf{Local}     & \textbf{prefix} \\
    \hline
    Eigen.  & [0.57,  \textbf{1.10}, 1.76](0.2)  & [0.67, \textbf{1.49}, 2.60](0.3)     & \texttt{01} \\
    Null.   & [-0.44, \textbf{-0.09}, 0.27](0.1) & [0.62, \textbf{1.82}, 4.74](0.5)     & \texttt{01} \\
    Subtle. & [-0.37, \textbf{0.31}, 6.12](1.4)  & [0.28, \textbf{25.72}, 197.00](15.8) & \texttt{0}  \\
    \hline
    \end{tabular}
    \label{tab: win margins}
    \end{center}
\end{table}

\paragraph{Distinguishability}
To compare effectiveness of the three schemes,
distinguishabilities, defined as follows, are 
presented in Table~\ref{tab: distinguishabilities}.


Worst-case distinguishability
\begin{align}
    D = \min \{\min(M_i)\}_{i > 0} - \max(M_0)
    \label{eq: worst-case distinguishability} ,
\end{align}
where $M_i$ is a marking factor of the $i$-winner(s) case.

Average-case distinguishability
\begin{align}
    d = \sqrt[K]{\Pi_{i > 0} (\bar{M}_i - \bar{M}_0)}
    \label{eq: worst-case distinguishability} ,
\end{align}
where $\bar{M}_i$ is an average marking factor of the $i$-winner(s) case
and $K = N - 1$.

\begin{table}[htbp]
    \caption{Distinguishability}
    \begin{center}
    \begin{tabular}{|l|c|c|c|c|}
    \hline
    \textbf{Scheme} & \multicolumn{4}{|c|}{\textbf{Distinguishability}} \\
    \cline{2-5}
                    & \multicolumn{2}{|c|}{\textbf{Worst-case}}   
                    & \multicolumn{2}{|c|}{\textbf{Average-case}} \\
    \cline{2-5}
                    & $D$ & $D/|\bar{M}_0|$ & $d$ & $d/|\bar{M}_0|$ \\
    \hline
    Eigen.  &  0.190 & 19.000 & 0.532 & 53.188\\
    Null.   &  0.220 & 0.524 & 0.548 & 1.306\\
    Subtle. &  0.550 & 0.753 & 1.140 & 1.561\\
    \hline
    \end{tabular}
    \label{tab: distinguishabilities}
    \end{center}
\end{table}



Subtle marking seems to provide better distinguishability for both worst and average cases in an absolute sense:
its corresponding $D$'s and $d$'s are the largest among ones of the three schemes.
However, relative values $D/|\bar{M}_0|$ and $d/|\bar{M}_0|$
reveal another perspective: eigenmarking seems to provide the best distinguishability:
the difference between the no-winner and some-winner is most noticeable 
when compared to the size of no-winner marking factor.

\section{Conclusion and discussion}
As shown in Table~\ref{tab: lambda0 vs others}, the marking factors in no-winner scenarios
have shown to be significantly different from ones in other scenarios in all marking schemes.
In addition, the winning margins --- Table~\ref{tab: win margins} showing how well the answers (if exist) will stand out ---
are also significantly large when using local versions.

Overall, all schemes seem to do well in both differentiation of non-winner case 
and finding the answers if exist.
Eigenmarking scheme is the only one having the non-zero global winning margin,
which means that the answers if exist always stand out regardless of prefix.
Comparing all three schemes, eigenmarking and subtle marking have shown to provided strong distinguishability,
with eigenmarking having the edge on the relative view.

However, the simplicity of the subtle marking is also beneficial in many aspects:
(1) using a fewer extra qubit (less prone to error and noise when used in a real hardware);
(2) easier analysis since there are only two sets of resulting amplitudes;
(3) less prone to noise (c.f. two previous schemes) since it uses full rotation ($\pi$ radian c.f. $\pi/2$).
and
(4) minimal change to the original Grover algorithm allows 
re-use of many techniques and analyses intended for Grover algorithm.

Our findings here have shown a great prospect on this eigenmarking approach.
The approach has potential much beyond being a crucial part in entailment checking.
It could have profound effects on development of quantum algorithms in general.
However, with this preliminary experiment on a two-qubit system,
there are still many issues that remain unanswered, including
scalability (how this mechanism works in a larger system),
formal analysis (how this mechanism works theoretically),
and applicability in a real quantum machine 
(how robust this mechanism is when run on a quantum computer in our current 
noisy intermediate-scale quantum era).

\emph{Code:}
\verb|https://github.com/tatpongkatanyukul/|
\verb|Publication/tree/main/QEigenMarking|



\vspace{12pt}

\section*{Appendix}

\subsection*{Examples of experimental results}
\begin{figure}[htbp]
    \includegraphics[width=0.48\textwidth]{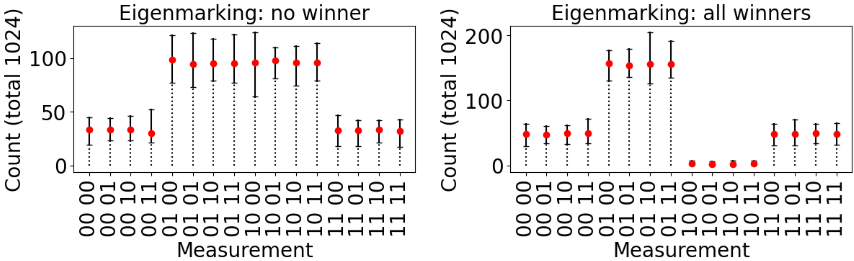}
\caption{Eigenmarking results: $\lambda = 0$ (no winner, left); $\lambda = 4/16$ (all winners, right). 
With additional two qubits, $\lambda = 4/16$ is the largest winning fraction.}
\label{fig: eigenmarking win 0 win 4}
\end{figure}

\begin{figure}[htbp]
    \includegraphics[width=0.48\textwidth]{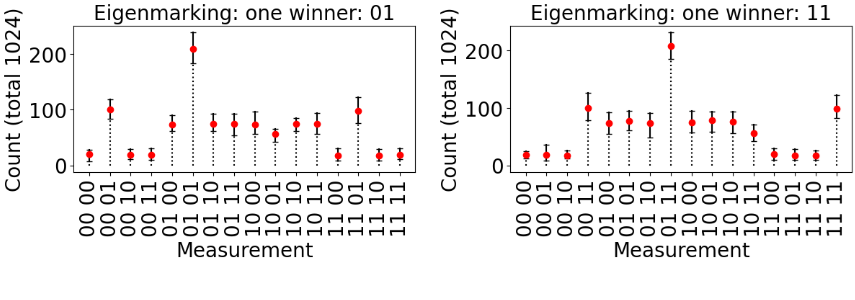}
\caption{Eigenmarking results: $\lambda = 1/16$ (one winner), arbitrarily selected.}
\label{fig: eigenmarking win 1}
\end{figure}

\begin{figure}[htbp]
    \includegraphics[width=0.48\textwidth]{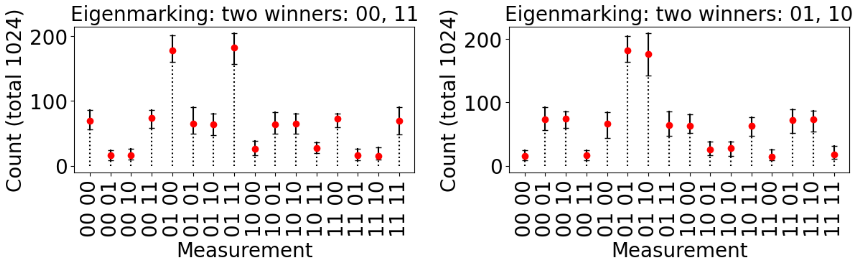}
    \caption{Eigenmarking results: $\lambda = 2/16$ (two winners), arbitrarily selected.}
\label{fig: eigenmarking win 2}
\end{figure}

\begin{figure}[htbp]
    \includegraphics[width=0.48\textwidth]{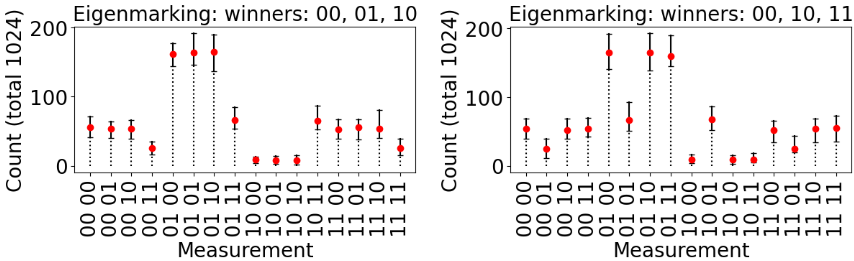} 
\caption{Eigenmarking results: $\lambda = 3/16$ (three winners), arbitrarily selected.}
\label{fig: eigenmarking win 3}
\end{figure}

\paragraph{Eigenmarking}
Fig.~\ref{fig: eigenmarking win 0 win 4} to \ref{fig: eigenmarking win 3} show the count results from 40 repeated experiments (each with 1024 runs) 
using eigenmarking.
The strength of this scheme is that the answers always have higher probabilities than other states, when there is at least one answer state.
The downside is that when there is no answer state, the indicator is not as obvious as ones of other schemes.

\begin{figure}[htbp]
    \includegraphics[width=0.48\textwidth]{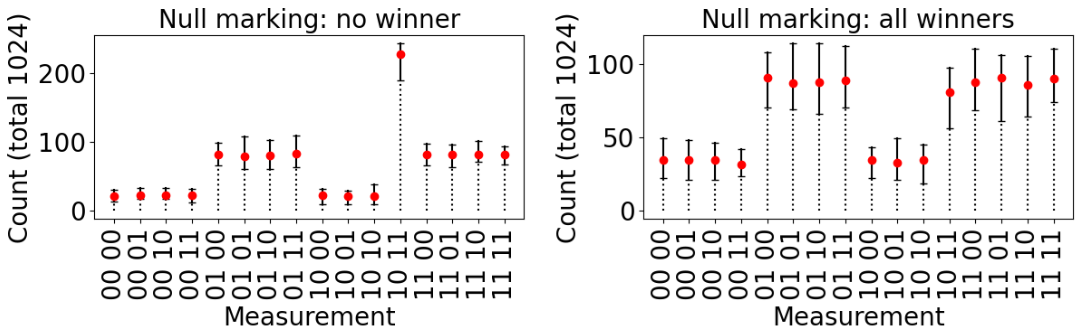}
\caption{Null marking results: $\lambda = 0$ (no winner, left); $\lambda = 4/16$ (all winners, right). 
With null marking, the state \texttt{1011} stands out when there is no winner.}
\label{fig: null marking win 0 win 4}
\end{figure}

\begin{figure}[htbp]
    \includegraphics[width=0.48\textwidth]{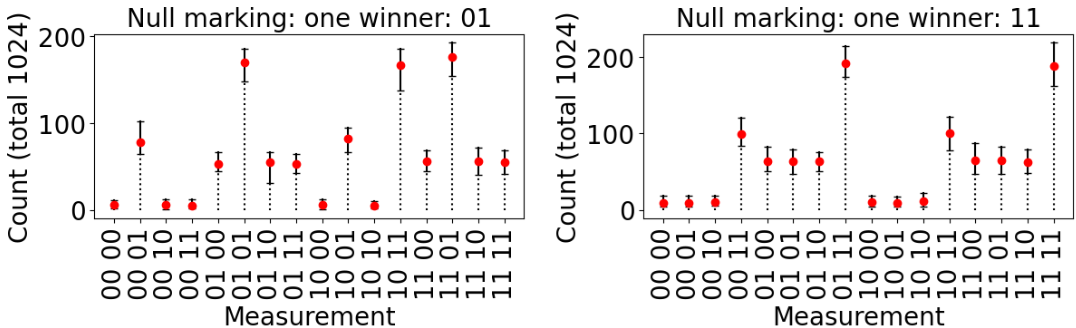}
\caption{Null marking results: $\lambda = 1/16$ (one winner), arbitrarily selected. 
With null marking, the state \texttt{1011} is suppressed when the \texttt{11} is a winning state,
but the answer (prefix \texttt{01}) still shows up fine.}
\label{fig: null marking win 1}
\end{figure}

\begin{figure}[htbp]
    \includegraphics[width=0.48\textwidth]{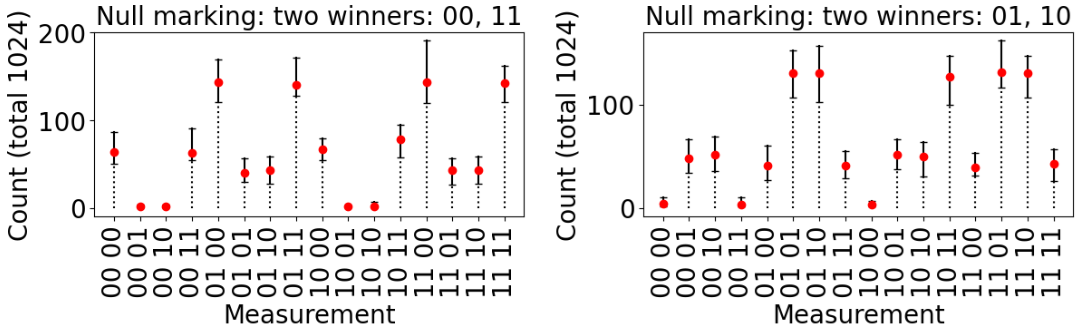}
\caption{Null marking results: $\lambda = 2/16$ (two winners), arbitrarily selected.}
\label{fig: null marking win 2}
\end{figure}

\begin{figure}[htbp]
    \includegraphics[width=0.48\textwidth]{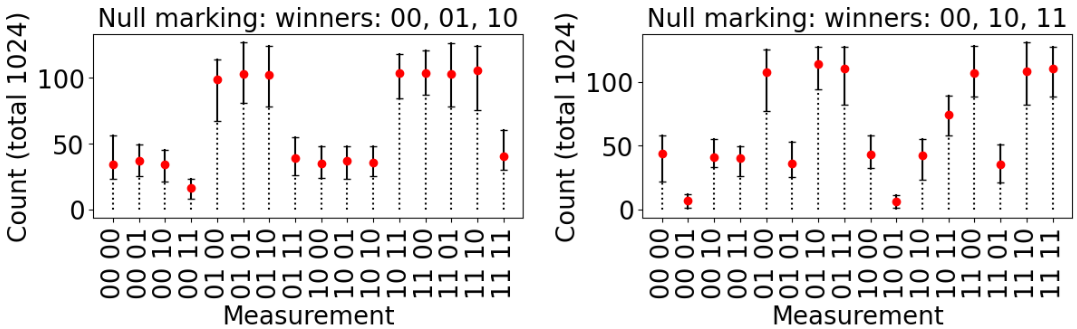}
\caption{Null marking results: $\lambda = 3/16$ (three winners), arbitrarily selected.}
\label{fig: null marking win 3}
\end{figure}

\paragraph{Null marking}
Fig.~\ref{fig: null marking win 0 win 4} to \ref{fig: null marking win 3} show the count results from 40 repeated experiments using eigenmarking with null marking.
Note that states with prefix \texttt{11} mirror the marked ones (prefix \texttt{01}).
These mirroring pairs could be utilized for complementary checking against noise. 

\begin{figure}[htbp]
    \includegraphics[width=0.48\textwidth]{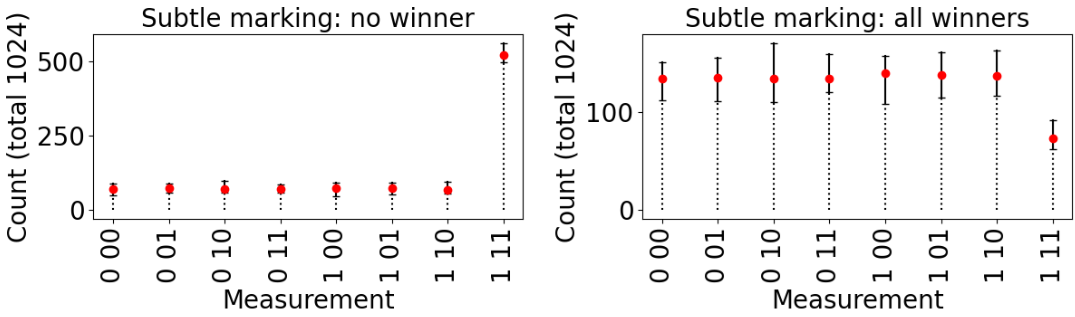}
\caption{Subtle marking results: $\lambda = 0$ (no winner, left); $\lambda = 4/16$ (all winners, right).}
\label{fig: subtle marking win 0 win 4}
\end{figure}

\begin{figure}[htbp]
    \includegraphics[width=0.48\textwidth]{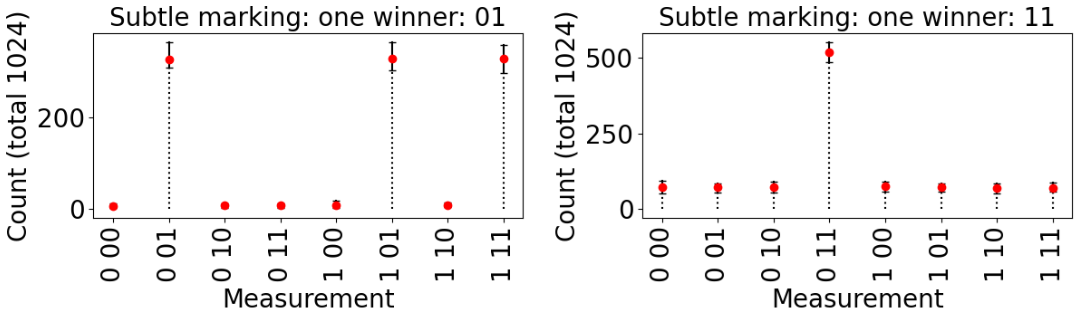}
\caption{Subtle marking results: $\lambda = 1/16$ (one winner), arbitrarily selected.}
\label{fig: subtle marking win 1}
\end{figure}

\begin{figure}[htbp]
    \includegraphics[width=0.48\textwidth]{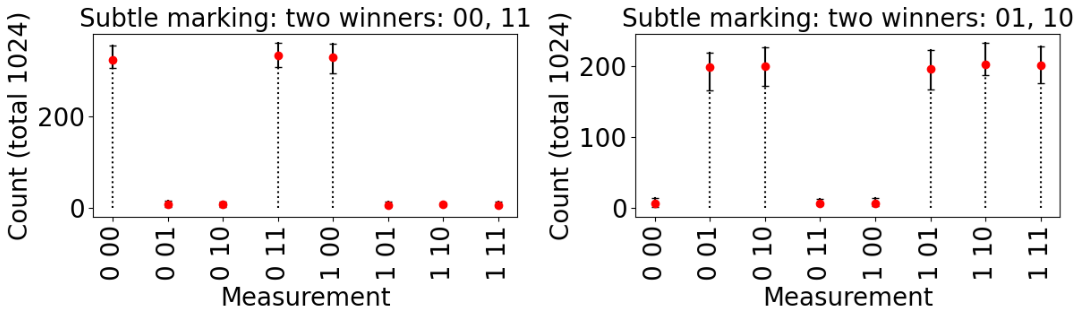}
\caption{Subtle marking results: $\lambda = 2/16$ (two winners), arbitrarily selected.}
\label{fig: subtle marking win 2}
\end{figure}

\begin{figure}[htbp]
    \includegraphics[width=0.48\textwidth]{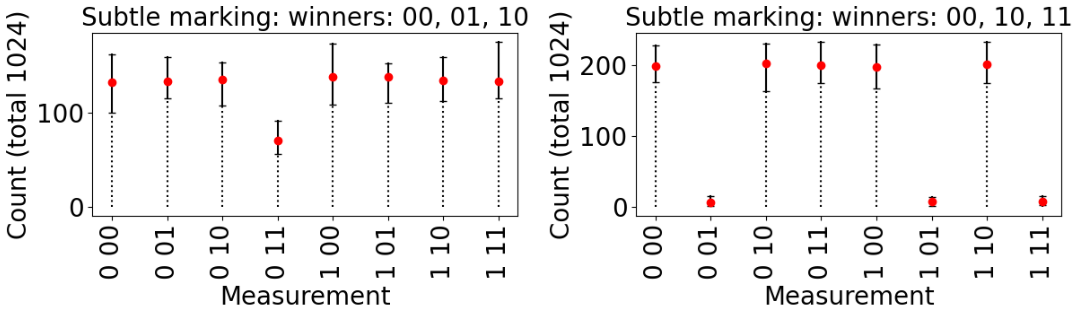}
\caption{Subtle marking results: $\lambda = 3/16$ (three winners), arbitrarily selected.}
\label{fig: subtle marking win 3}
\end{figure}

\paragraph{Subtle marking}
Fig.~\ref{fig: subtle marking win 0 win 4} to \ref{fig: subtle marking win 3} show the count results from 40 repeated experiments using eigenmarking with subtle marking.

\subsection*{Closed form formulation of Grover iteration}
The closed forms in equations~\ref{eq: kj closed} and~\ref{eq: ellj closed}
can be derived from equations~\ref{eq: kj} and~\ref{eq: lj} as follows.

Let's consider $j = 0$.
\begin{align}
k_1    &= \frac{N-2}{N} \underbrace{\frac{1}{\sqrt{N}}}_{k_0} + \frac{2 (N-1)}{N} \underbrace{\frac{1}{\sqrt{N}}}_{\ell_0}
\label{eq: k1} , \\
\ell_1 &= \frac{N-2}{N} \frac{1}{\sqrt{N}} - \frac{2}{N} \frac{1}{\sqrt{N}}
\label{eq: l1} .
\end{align}

For $\sin(\theta) \equiv \frac{1}{\sqrt{N}}$, we can derived $\cos(\theta) = \frac{\sqrt{N-1}}{\sqrt{N}}$ 
using Pythagorean theorem.
That is,
\begin{align}
k_1    &= \frac{N-2}{N} \sin(\theta) + \frac{2 (N-1)}{N} \cos(\theta)
\label{eq: k1} , \\
\ell_1 &= \frac{N-2}{N} \frac{1}{\sqrt{N}} - \frac{2}{N} \frac{1}{\sqrt{N}}
\label{eq: l1} .
\end{align}

From trigonometry, $\cos(2 \theta) = 1 - 2 \sin^2(\theta)$,
we have $ 1 - \frac{2}{N} = \frac{N-2}{N}$
and from Pythagorean theorem, we also have $\sin(2 \theta) = \frac{2 \sqrt{N - 1}}{N}$.
Hence,
\begin{align}
k_1 &= \cos(2 \theta) \sin(\theta) + \sin(2 \theta) \cos(\theta)
\label{eq: k1 cos sin} , \\
\ell_1 &= \frac{\sqrt{N-1}}{\sqrt{N-1}} \left( \frac{N-2}{N} \frac{1}{\sqrt{N}} - \frac{2}{N} \frac{1}{\sqrt{N}} \right)
\nonumber \\
&= \frac{1}{\sqrt{N-1}} \left( \frac{N-2}{N} \frac{\sqrt{N-1}}{\sqrt{N}}
 - \frac{2}{N} \frac{\sqrt{N-1}}{\sqrt{N}} \right)
\nonumber \\
&= \frac{1}{\sqrt{N-1}} \left( \cos(2 \theta) \cos(\theta) - \sin(2 \theta) \sin( \theta ) \right)
\label{eq: l1 cos sin} .
\end{align}

From trigonometry identities, $\sin(A + B) = \cos(A) \sin(B) + \sin(A) \cos(B)$
and $\cos(A + B) = \cos(A) \cos(B) - \sin(A) \sin(B)$.
We get,
\begin{align}
k_1 &= \sin(3 \theta)
\label{eq: k1 closed} , \\
\ell_1 &= \frac{1}{\sqrt{N-1}} \cos(3 \theta)
\label{eq: l1 closed}.
\end{align}

Both equations agree with equations~\ref{eq: kj closed} and~\ref{eq: ellj closed}.
Let's turn to a general case.
Suppose equations~\ref{eq: kj closed} and~\ref{eq: ellj closed} hold,
Iteration $j-1$ has,
$k_{j-1} = \sin (\phi)$
and 
$\ell_{j-1} = \frac{1}{\sqrt{N -1}} \cos (\phi)$
where 
$\phi =  (2(j-1) + 1) \theta$ $= (2j - 1) \theta$.

Hence,
\begin{align}
k_j &= \frac{N - 2}{N} \sin (\phi) + \frac{2(N - 1)}{N} \frac{\cos (\phi) }{\sqrt{N -1}} 
\nonumber , \\
\ell_j &= - \frac{2}{N} \sin (\phi) + \left(\frac{N - 2}{N}\right) \frac{\cos (\phi) }{\sqrt{N -1}} 
\nonumber .
\end{align}

Similarly, we can write
\begin{align}
k_j &= \cos(2 \theta) \sin (\phi) + \sin(2 \theta) \cos (\phi) 
\nonumber \\
&= \sin (2 \theta + \phi) = \sin (2 \theta + (2j - 1) \theta) = \sin((2j+1) \theta)
\nonumber , \\
\ell_j &= \frac{1}{\sqrt{N -1}} \left( \right. 
- \frac{2 \sqrt{N -1}}{N} \sin (\phi) 
\nonumber \\
&\; + \left(\frac{N - 2}{N}\right) \frac{\sqrt{N -1}}{\sqrt{N -1}} \cos (\phi) \left.\right) 
\nonumber \\
&= \frac{1}{\sqrt{N -1}} \left(\right. - \sin(2 \theta ) \sin (\phi)
\nonumber \\
&\; + \cos(2 \theta) \cos (\phi) \left.\right)
\nonumber \\
&= \frac{1}{\sqrt{N -1}} \cos( 2 \theta + \phi) = 
\frac{1}{\sqrt{N -1}} \cos((2j+1) \theta)
\nonumber .
\end{align}
Both equations agree with the closed forms.
Therefore, the closed forms (\ref{eq: kj closed} and~\ref{eq: ellj closed}) are valid.

\end{document}